\documentclass{elsart}

\usepackage{natbib}
\usepackage{psfig}
\usepackage{graphicx}
\usepackage{epsfig}

\usepackage{amssymb}
\def\ut#1{\mathop{\vtop{\ialign{##\crcr
     $\hfil\displaystyle{#1}\hfil$\crcr\noalign
     {\kern1pt\nointerlineskip}\hbox{$\hfil\sim\hfil$}\crcr
     \noalign{\kern1pt}}}}}

\def\undersymbol#1#2{\mathop{\vtop{\ialign{##\crcr
     $\hfil\displaystyle{#2}\hfil$\crcr\noalign
     {\kern1pt\nointerlineskip}\hbox{$\hfil#1\hfil$}\crcr
     \noalign{\kern1pt}}}}}
\def\arcsec{^{\prime\prime}}

\begin{document}

\begin{frontmatter}


\title{XMM-Newton observation of MACHO 104.20906.960: a dwarf nova candidate with a 2\,h period}
\author[esa]{A.A. Nucita\corauthref{corresponding1}},
\ead{anucita@sciops.esa.int}
\corauth[corresponding1]{Corresponding author: A.Nucita}
\author[esa]{S. Carpano},
\author[dipfisicaunile]{F. De Paolis},
\author[dipfisicaunile]{G. Ingrosso},
\author[dipfisicaunile]{B.M.T. Maiolo}, and
\author[esa]{M. Guainazzi}
\address[esa]{XMM-Newton Science Operations Centre, ESAC, ESA, PO Box 78, 28691 Villanueva de la $\rm Ca\tilde{n}ada$, Madrid, Spain}
\address[dipfisicaunile]{Dipartimento di
Fisica, Universit\`a del Salento, and {\it INFN}, Sezione di Lecce, CP
193, I-73100 Lecce, Italy}

\begin{abstract}
The binaries known as cataclysmic variables are particular binary systems in which the primary star (a white dwarf) accretes material from a secondary via Roche-lobe mechanism. Usually, these objects have orbital period of a few hours so that a detailed  temporal analysis can be performed. Here, we present Chandra and ${\it XMM}$-Newton observations of a dwarf nova candidate identified in the past by optical observations towards the galactic Bulge and labeled as MACHO 104.20906.960. After a spectral analysis, we used the Lomb-Scargle technique for the period search and evaluated the confidence level using Monte-Carlo simulations.
In this case, we found that the $X$-ray source shows a period of $2.03_{-0.07}^{+0.09}\,$ hours (3$\sigma$ error) so that it is most likely a system of interacting objects. The  modulation of the signal was found with a confidence level of $>$99\%. The spectrum can be described by a two thermal plasma components with X-ray flux in the 0.3--10\, keV energy band of $1.77_{-0.19}^{+0.16}\times10^{-13}$ erg s$^{-1}$ cm$^{-2}$. We find that the distance of the source is approximately 1 kpc thus corresponding to a luminosity $L_{X}\simeq 2\times 10^{31}$ erg s$^{-1}$.
\end{abstract}

\end{frontmatter}



\section{Introduction}

The observation and subsequent identification of some $X$-ray sources as low mass $X$-ray binaries (LMXRBs), high mass $X$-ray binaries (HMXRBs) and cataclysmic variables (CVs) plays a crucial role in the study of collapsed objects since the physics that they involve is strictly related with the strong gravity regime.

As an example, CVs are binary systems constituted by a white dwarf primary star interacting with a secondary object whose spectral type may range from G to M classes. The secondary star funnels matter on the primary via the formation of a Roche-lobe. Moreover, how the accretion on the white dwarf occurs strongly depends on some key parameters as the magnetic field strength. In particular, a disk surrounding the white dwarf will form if the primary star presents a low magnetic field (as it happens in the dwarf nova systems), while a mass stream appears in highly magnetized binaries as the so called polars. In the dwarf novae sample, the accreting material may trigger some {\it outburst} episodes on timescales ranging from weeks to months so that the typical brightness of the system increases of a few magnitudes (see e.g. \citealt{narayan1993}, and references therein). A detailed review of the CV properties can be found in \citet{erik}.

In the last decade, different micro-lensing programs as OGLE and
MACHO (see e.g. \citealt{udalski1997}, and \citealt{alcock2001}) monitored several directions in the sky in order to collect data and give proof of the existence of dark matter within the Galaxy. These observational campaigns allowed also to identify several eruptive variables as CVs and many authors compiled detailed catalogues with the properties of the observed sources (see e.g. \citealt{cielinski}, for the CVs identified in the MACHO database).

The cataclysmic variables which are part of the dwarf novae sample are usually discovered looking for several observational features in the optical as well as in the UV band. However, it has been noted that the efficiency in the detection depends strongly on the amplitude of the outbursts (and their separation in time) so that observations in bands different than the optical and the ultraviolet ones are required to confirm the nature of these sources. Hence, the $X$-ray emission coming from these source represents a powerful tool of investigation despite of the relatively low luminosity ($\simeq 10^{31}-10^{33}$ erg s$^{-1}$, see e.g. \citealt{lamb82} and \citealt{erik}) of dwarf novae. In this respect, {\it XMM}-Newton (\citealt{jansen2001}) thanks to its large effective area is suitable for this kind of search (see e.g.  \cite{ramsey2001a,ramsey2001b} for a detailed analysis of the first {\it XMM}-Newton observation of a cataclysmic variable system).

Here, we report on a $\sim 100$ ks {\it XMM}-Newton observation of a
dwarf nova candidate towards the galactic Bulge. The target
name, as given in the MACHO database, is MACHO 104.20906.960 and a
detailed description of the source optical properties can be found
in \cite{cielinski}. The source observed by $XMM$-Newton shows a
light curve with a strong periodic behavior (with a period of
$\simeq 2.03\,$ h). We estimate the $X$-ray flux of the
observed dwarf nova to be $1.77_{-0.19}^{+0.16}\times10^{-13}$ erg s$^{-1}$ cm$^{-2}$  in the 0.3-10 keV.
The paper is structured as follows: in Sect. 2, we
briefly report about the optical properties of the dwarf nova
candidate. In Sect. 3 and Sect. 4, we give details on the data
reduction and on the spectral and timing analysis conducted on
MACHO 104.20906.960, while in Sect. 5 we draw some conclusions aiming that a cross correlation search in different bands of this kind of sources is performed in the future.

\section{The optical properties of the dwarf nova candidate MACHO 104.20906.960}

The MACHO project (see \citealt{alcock1997} and  \citealt{alcock1999}) collected images of millions
of stars in the direction of the galactic Bulge and Small and Large Magellanic Clouds to detect
micro-lensing events and trace the amount of baryonic dark matter distributed in the Galaxy.

Because of the large amount of data collected in many years, thousand of eruptive variables as well as CVs were identified so that it was possible to compile rich catalogues with the main properties of each of the observed source. In particular, the discovery of $28$ new dwarf nova candidates was the main result of \cite{cielinski}, to which we refer for more details about the characteristics of each of the newly discovered CV systems.

The result of the analysis of \cite{cielinski} is listed in their Table 1, and the light curves of all the identified dwarf nova candidates are given in their Fig. 1.

\begin{figure}[htbp]
\vspace{6cm} \includegraphics{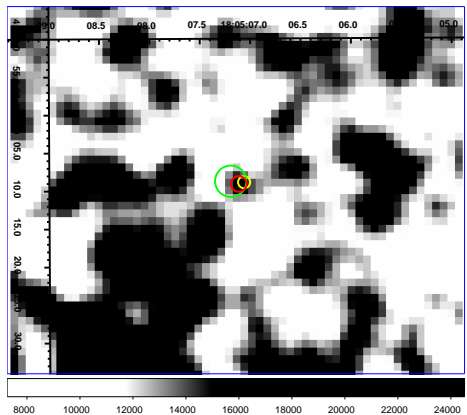}
\caption{The DSS field of view (red band) around MACHO 104.20906.960 ($60\arcsec\times60\arcsec$) is given with the green ($2\arcsec$), red ($1\arcsec$) and yellow ($1\arcsec$) rings centered on the $XMM$-Newton, MACHO catalogue and Chandra source coordinates, respectively.}
\label{f7}
\end{figure}

In the present work, we study the dwarf nova candidate labeled
as MACHO 104.20906.960 that was serendedipitously observed in 2005. The coordinates of the target are
$\alpha_{\rm J2000}=18^{\rm h}05^{\rm m} 07^{s}12$ and $\delta_{\rm J2000}=-27^\circ 43' 09\arcsec 13$. In the following, we briefly report the main results of the analysis conducted by \cite{cielinski}. In particular, they found that the dwarf nova MACHO 104.20906.960 has a light curve with length $2412$ days during which 4 outbursts were possibly observed. The source has a typical $V$ magnitude far from the outbursts of $17.6$ and a median color
index $(V-R)=0.69$. The magnitude at the maximum of the outburst
increases of $\Delta V=1.5$ (the outburst color amplitude being
$\Delta (V-R)=-0.3$). In Fig. \ref{f7}, for completeness we give a DSS field of
view (red band) around MACHO 104.20906.960 ($60\arcsec\times60\arcsec$).
Here, the red ring is centered on the MACHO database coordinates and has a radius of $1\arcsec$ corresponding to the astrometric precision achieved by the MACHO experiment. In the same figure we also give a green ring (with radius of $2\arcsec$) and yellow ring (with radius of $1\arcsec$) centered on the source identified by {\it XMM}-Newton and Chandra, respectively (for further details see Section 3).
As clearly stated by \cite{cielinski}, when the catalogue of the dwarf novae candidates was compiled, it was not possible to find a $X$-ray counterpart of any of the newly discovered CVs. In fact, when the authors assumed a $F_{X}$ to $F_{opt}$ ratio lower than 0.5, they expected to have all the sources below the detection limit ($F_{\rm{0.5-2.5 keV}}\simeq 10^{-13}$ erg s$^{-1}$ cm$^{-2}$) of the ROSAT/PSPC observations. Hence, as the same authors suggested (\citealt{cielinski}), $X$-ray observations of dwarf novae necessarily require more sensitive instruments as the CHANDRA and {\it XMM}-Newton satellite.

For this reason, in the next Section, we will consider a deep {\it XMM}-Newton observation towards the identified dwarf nova candidate MACHO 104.20906.960 and use the Chandra satellite and its better astrometric precision to confirm its nature as predicted by \cite{cielinski}.

\section{{\it XMM}-Newton and Chandra observation of the dwarf nova candidate MACHO 104.20906.960}
\begin{figure*}[htbp]
\vspace{0.8cm}
\begin{center}
$\begin{array}{c@{\hspace{0.1in}}c@{\hspace{0.1in}}c}
\epsfxsize=2.8in \epsfysize=2.8in \epsffile{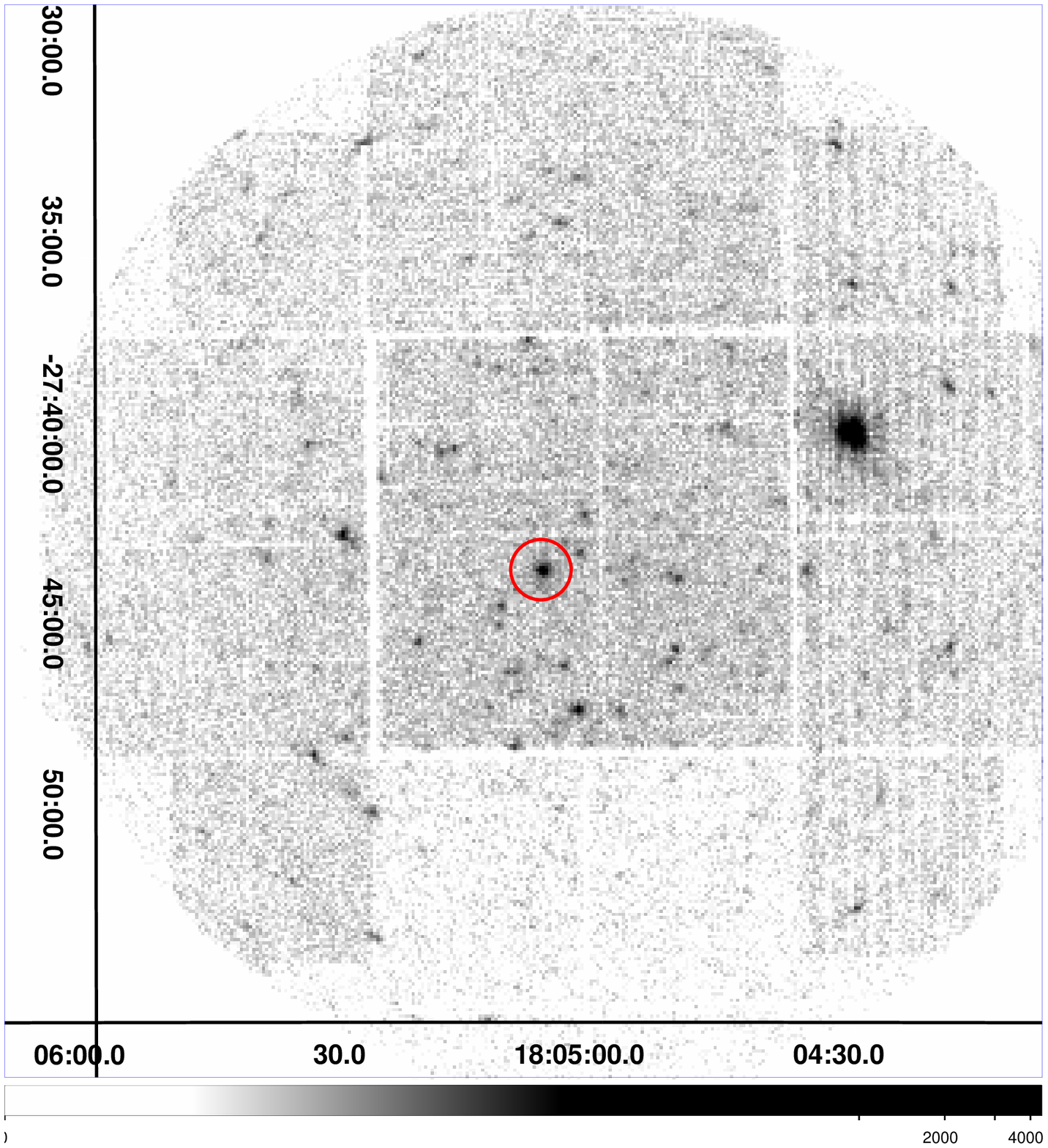} &
\epsfxsize=2.8in \epsfysize=2.5in \epsffile{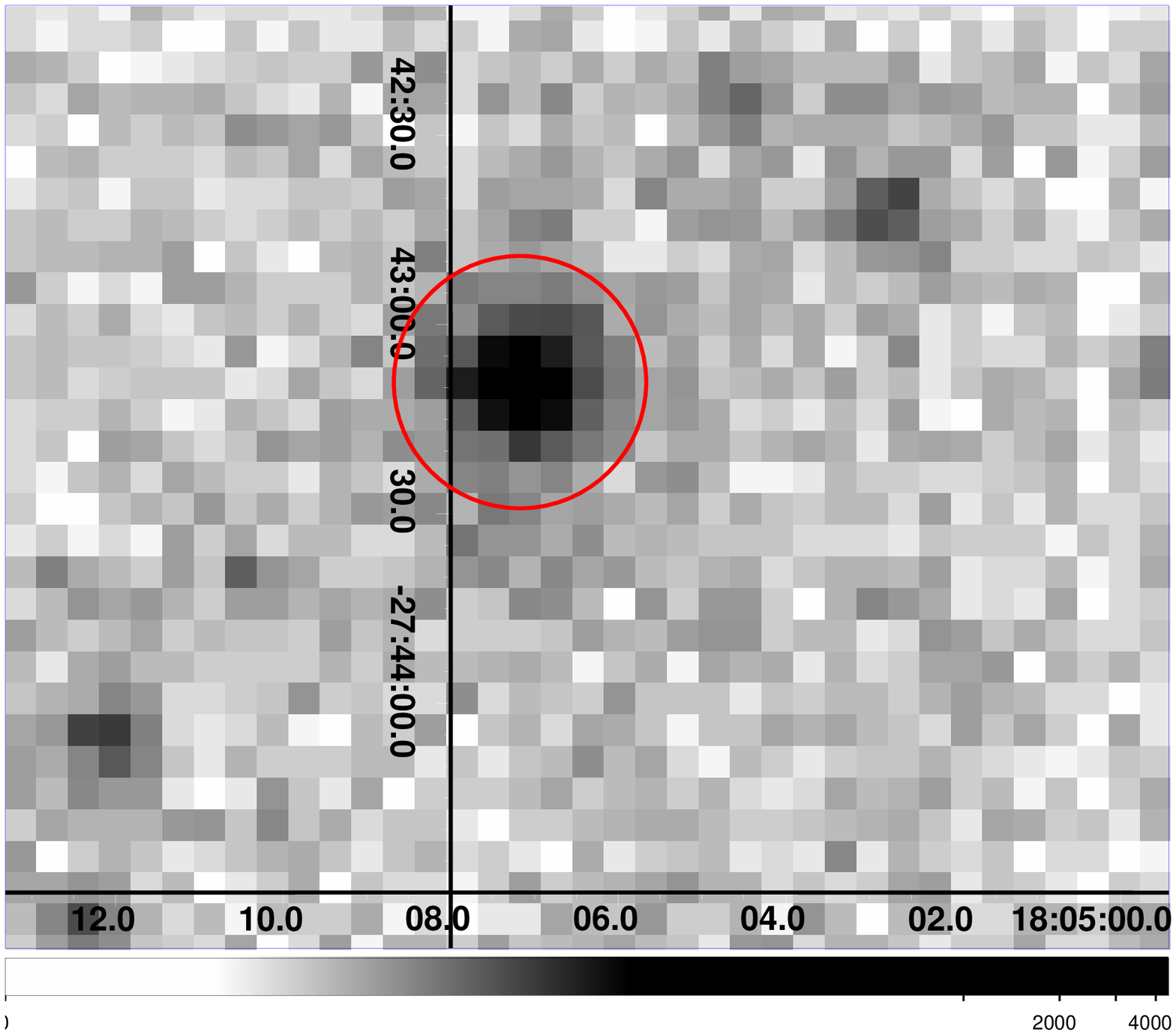}  \\
\end{array}$
\end{center}
\caption{A combined image (in the band 0.2-10 keV) of the MOS 1, MOS 2 and PN cameras of the full {\it XMM}-Newton field of view (left panel) is shown. On the right panel, the dwarf nova candidate is indicated by the red circle (with radius $\simeq 20\arcsec$).}
\label{f1}
\end{figure*}

Here, we are reporting about a {\it XMM} observation conducted
in the direction of the Galactic Bulge and the serendipitous discovery of the X-ray counterpart of the dwarf nova candidate MACHO 104.20906.960. The field of view around the source was observed by {\it XMM}-Newton in two occasions: in September 2005 (Observation ID 30480) with both
the EPIC MOS and PN cameras (\citealt{mos,pn}) operating in the medium filter mode (exposure time of $\simeq 10$ ks) and in October 2005 (Observation ID 30597) with the thin filter and
for a scheduled observational time of $\simeq 100$ ks. Unfortunately, in the first observation, the source is close to the camera borders so that, in the following, we do not consider it in our analysis.

The EPIC data files (ODFs) were processed using the
{\it XMM}-Science Analysis System (SAS version $7.0.0$). The raw data were processed using the
latest available calibration constituent files. We ran the {\it emchain} and {\it
epchain} tools and obtained several event list files on which the usual screening
criteria were applied. Thus, after we searched and rejected periods affected by soft protons flares the remaining good time intervals resulted in exposures with effective time of $\simeq 96$ ks, $\simeq 98$ ks, and $\simeq 94$ ks for
MOS 1, MOS 2, and PN, respectively.

By using the {\it edetect\_chain} tool, the coordinates of the source MACHO 104.20906.960 were determined to be $\alpha_{\rm J2000}=18^{\rm h}05^{\rm m} 07^{s}20$ and $\delta_{\rm J2000}=-27^\circ 43' 08\arcsec 76$.
In this respect, note that a systematic shift between the X-ray and optical position of the sources of $-2.52\arcsec$ and $-3.09\arcsec$ (at a level of $1\sigma$) exists in right ascension for MOS 1 and MOS 2. The shift on declination in the same two cameras is $-1.19\arcsec$ and $0.41\arcsec$, respectively. For this reason, in the following we assume that the error associated to the two celestial coordinates is at least $\sim 2\arcsec$.

In Fig. \ref{f1}, we give a combined image (in the band 0.2-10 keV) of the MOS 1, MOS 2 and PN cameras of the full {\it XMM}-Newton field of view (left panel). The deep exposure has shown the existence of several new $X$-ray sources\footnote{This $XMM$-Newton field of view was also particular interesting because of the discovery of the long parallax microlensing event MACHO-96-BLG-5 (see \citealt{poinsexter2005} for a detailed description of the event parameters, and \citealt{nucita2006} for an $X$-ray search for the lens).} (as the bright source at the up-right corner which was recently identified as a M-type X-ray binary with a 494\,s-pulse period neutron star, \citealt{nucita2007}) which will be investigated elsewhere. On the right panel, the dwarf nova candidate studied in the present paper is indicated by the red circle ($\simeq 20\arcsec$).

The {\it Chandra} satellite has a better angular resolution ($\simeq 1\arcsec$, see \citealt{chandra} for details) than the $XMM$-Newton telescope and therefore it is worth using {\it Chandra} to improve the X-ray astrometry and remove any doubt that the source studied here is the $X$-ray counterpart of the dwarf nova found in the MACHO catalogue. The Chandra satellite observed toward the target in 2003 (Observation ID 3789) with the ACIS-S camera and for $\simeq 10$ ks . We created images in the full 0.3-8 keV band and search for discrete sources by the CIAO {\it celldetect} tool with a threshold signal-to-noise detection value of $3$. This resulted in the identification of a source (with signal-to-noise ratio of $\simeq 4$) with coordinates $\alpha_{\rm J2000}=18^{\rm h}05^{\rm m} 07^{s} 067$ and $\delta_{\rm J2000}=-27^\circ 43' 08\arcsec 90$ and astrometric precision of $\simeq 1\arcsec$.

The green and  yellow rings in Fig. \ref{f7} are centered on the {\it XMM}-Newton and Chandra source coordinates and have a radius corresponding to the astrometric precision of each satellite, respectively. Hence, we believe that the $XMM$-Newton identified source corresponds to the $X$-ray counterpart of the dwarf nova candidate MACHO 104.20906.960 detected by \cite{cielinski}.

\subsection{Spectral analysis of MACHO 104.20906.960}

The source spectra were extracted in a circular region centered on the nominal position of the target in the
three EPIC cameras (extraction circles with radius $\simeq 30\arcsec$), while the background spectra were accumulated in circular regions on the same chip and at the same vertical location. The resulting spectra were rebinned to have at least 25 counts per energy bin.

The spectra were simultaneously fitted with XSPEC (version 12.4.0).
\begin{figure*}[htbp]
\vspace{8.5cm} \includegraphics{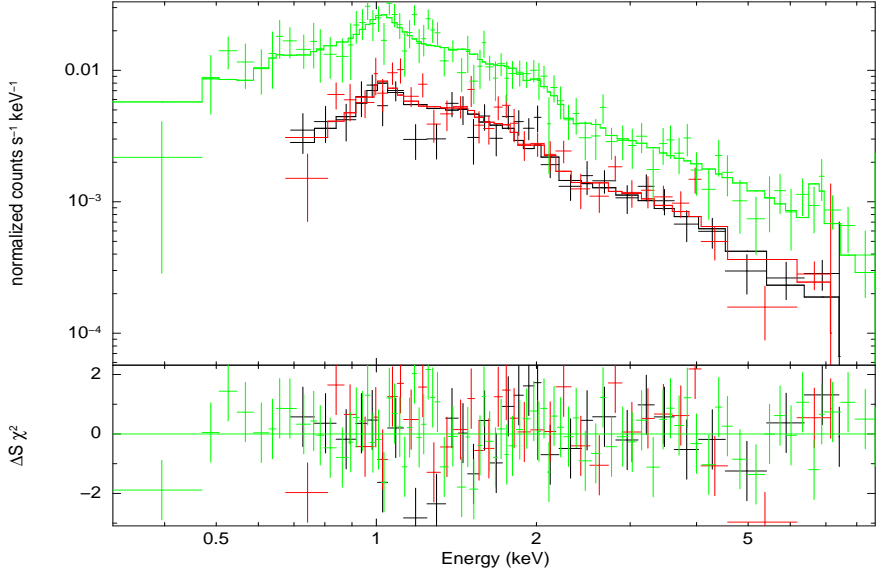}
\caption{The best-fitting model (see text for details) is given together with the MOS 1, MOS 2 and PN data.}
\label{f2}
\end{figure*}
A single temperature thermal plasma model with absorption by neutral gas (MEKAL and WABS models in XSPEC) gave a fit with $\chi^2_{\nu}=1.2$ (144 d.o.f.) with temperature $kT=6.43_{-1.17}^{+1.06}$ keV and a hydrogen column density $N_H=(1.6 \pm 0.3)\times 10^{21}$ cm$^{-2}$. The normalization of this component is $N=(8.3\pm 0.4)\times 10^{-5}$. Adding a second thermal plasma model yielded a significantly better fit ($\chi^2_{\nu}=1.00$, for 142 d.o.f.) with values of the temperature $kT_1=1.27_{-0.32}^{+0.13}$ keV and $kT_2=10.65_{-3.40}^{+4.20}$ keV. The fitted hydrogen column density\footnote{Note that the column density obtained from the fit procedure is consistently lower than that estimated to be in our Galaxy ($N_H\simeq 3.3\times 10^{21}$ cm$^{-2}$, \citealt{dickey}).} is $N_H=(1.8_{-0.3}^{+0.3})\times 10^{21}$ cm$^{-2}$. The normalizations corresponding to the two components are
$N_1=(9.1_{-3.8}^{+4.5})\times 10^{-6}$ and $N_2=(7.5_{-0.5}^{+0.6})\times 10^{-5}$, respectively.
The addition of a third thermal plasma component or a blackbody one did not improve the fit in a significant way.
In Fig. \ref{f2}, we show the MOS 1, MOS 2, and PN spectra ($0.3-10$ keV) for the source and the respective best fits. The absorbed flux in the 0.3-10 keV is $F^{\rm{Abs}}_{\rm{0.3-10}}=1.48_{-0.17}^{+0.14}\times10^{-13}$ erg s$^{-1}$ cm$^{-2}$ corresponding to a corrected (unabsorbed) flux (in the same band) of $F^{\rm{Cor}}_{\rm{0.3-10}}=1.77_{-0.19}^{+0.16}\times10^{-13}$ erg s$^{-1}$ cm$^{-2}$ (all the errors are quoted at the $90\%$ confidence level)

The source count rate detected by the Chandra satellite during the 2003 observation corresponded to $\simeq (4.04\pm0.02)\times 10^{-3}$ counts/sec. In spite of the fact that the count rate is slightly low, we extracted the spectrum from a circular aperture centered on the source and, once imported within XSPEC, we fitted it by using the same best fit model obtained with the {\it XMM}-Newton data. After extrapolating the flux in the $0.3-10$ keV band we got that the Chandra absorbed flux is $F^{\rm{Abs}}_{0.3-10 {\rm keV}}\ut< 2.9\times 10^{-13}$ erg s$^{-1}$ cm$^{-2}$, i.e. consistent with the result obtained by using {\it XMM}-Newton and quoted above.

\subsection{Timing analysis of MACHO 104.20906.960}

We extracted the light curves of the source from all the three EPIC cameras using apertures $\simeq 30\arcsec$ in radius and centered on the MACHO 104.20906.960 position. The background light curves were extracted from the same CCD where the source was detected, scaled and subtracted from the source light curves.
To increase the signal to noise ratio, the counts from the MOS 1, MOS 2 and PN cameras were summed. We repeated the entire process for the energy range 0.2-10 keV (to search for any periodicity in the full light curve) and for the ranges $S=0.2-1$ keV and $H=1-10$ keV to evaluate the hardness ratio light curve in the two above energy bands and search for any change in the source state (see next).

\begin{figure}[htbp]
\vspace{6cm} \includegraphics{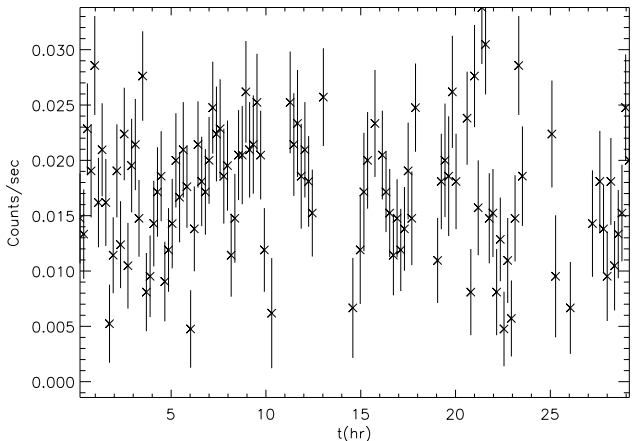}
\caption{The 0.2-10 keV light curve of the 100 ks observation of MACHO 104.20906.960, as binned at $700$ s. Here the  MOS 1, MOS 2 and PN counts are summed and background corrected.}
\label{f5}
\end{figure}

\begin{figure}[htbp]
\vspace{6cm} \includegraphics{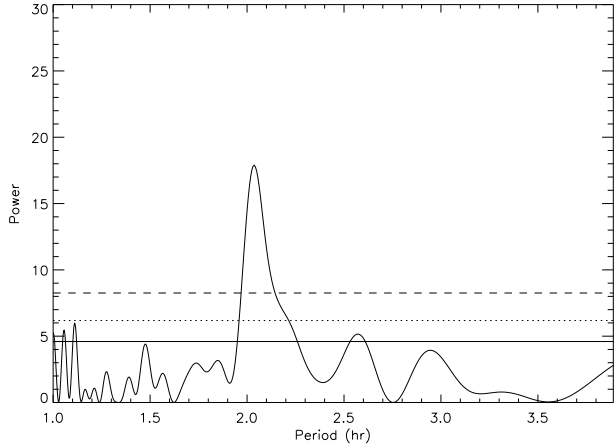}
\caption{Search for periodicities in the {\it XMM}-Newton light curve of MACHO 104.20906.960 using a
Lomb-Scargle periodogram analysis. The full, dotted, and dashed lines represent the 68\%, 90\%, and
99\% confidence levels, respectively.}
\label{f3}
\end{figure}

\begin{figure}[htbp]
\vspace{6cm} \includegraphics{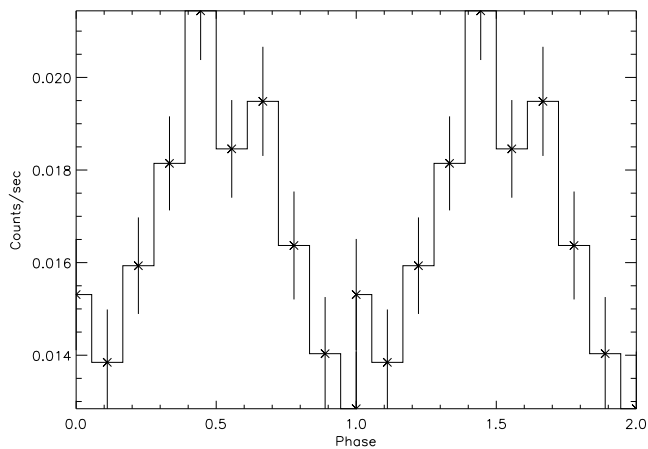}
\caption{The 0.2--10\,keV light curve folded at $2.03$\,h using 10 bins. Phase zero is associated to the
beginning of the {\it XMM}-Newton observation.}
\label{f4}
\end{figure}

The 0.2-10 keV light curve of MACHO 104.20906.960 during the 100 ks observation is given in Fig. \ref{f5} as binned at $700\,$s. We searched for periodicity between 20 sec and $10\,$h, and found a periodic signal at $2.03\,$h using a Lomb-Scargle periodogram (\citealt{Lomb1976,Scargle1982}). The confidence level of our detection was estimated with Monte Carlo simulations assuming a null hypothesis of white noise, and the results are plotted in Fig.~\ref{f3}, with the 68\%, 90\%, and 99\% confidence levels represented by the full, dotted and dashed horizontal lines, respectively.
We found that the $2.03\,$h period is significant at a confidence level $>99\%$
(see \citealt{Carpano2007} for more details on the applied technique). Thus, at 3$\sigma$ confidence level, the detected period of the observed source is $2.03_{-0.07}^{+0.09}\,$ h.
Assuming that the $X$-ray emission comes from the boundary layer (where the $X$-ray emission is produced, see \citealt{erik} for a review), the period estimated above is consistent with the typical dwarf nova orbital period. The XMM light curve folded at the detected period of $2.03\,$h is shown in  Fig.~\ref{f4}.


\begin{figure}[htbp]
\vspace{6cm} \includegraphics{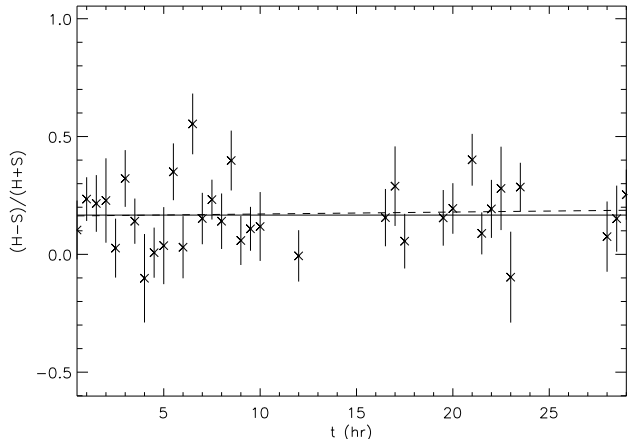}
\caption{The hardness ratio $HR=(H-S)/(H+S)$ light curve (binned at $1800$ s) of MACHO 104.20906.960. The used energy bands are $S=0.2-1$ keV and $H=1-10$ keV, respectively.}
\label{f6}
\end{figure}

Using the light curve extracted in the soft ({\it S}) and hard ({\it H}) parts of the spectrum (as described above), we defined the usual hardness ratio $HR=(H-S)/(H+S)$ and built the corresponding hardness ratio light curve to search for variability in the state of the source. The hardness ratio light curve is given in Fig. \ref{f6}. Here, the solid horizontal line represents the value of the average hardness ratio, while the almost superimposed dashed line is the result of a linear fit to the binned data, clearly showing that the state of the source remains constant (i.e. without overall significant changes) during the observation, as it would be expected since we observed the source during its quiescent state (i.e. far from an outburst).

\section{Discussion and conclusions}

In the present work, we studied the $X$-ray counterpart of the dwarf nova candidate labeled
as MACHO 104.20906.960.

A two temperature thermal plasma model with absorption by neutral gas (MEKAL and WABS models in XSPEC) gave an acceptable fit with values of the temperature $kT_1=1.27_{-0.32}^{+0.13}$ keV and $kT_2=10.65_{-3.40}^{+4.20}$ keV. The fitted hydrogen column density is $N_H=(1.8\pm 0.3)\times 10^{21}$ cm$^{-2}$ cm$^{-2}$ and the corrected (unabsorbed) flux resulted to be $F^{\rm{Cor}}_{0.3-10}=1.77_{-0.19}^{+0.16}\times10^{-13}$ erg cm$^{-2}$ s$^{-1}$, all the errors being at the $90\%$ confidence level. We found a periodicity of $2.03_{-0.07}^{+0.09}\,$ h in the source light curve by using a Lomb-Scargle periodogram.

If the $X$-ray measured hydrogen column density ($N_H\simeq 1.8\times 10^{21}$ cm$^{-2}$) is entirely due to interstellar extinction, then using the standard $N_H-A_v$ relation (see e.g. \citealt{gorenstein}), one has $A_v\simeq 0.9$, being this un upper limit to the extinction since many dwarf novae show evidence for intrinsic $X$-ray absorbtion. Hence, we estimate the unabsorbed $V$ magnitude of the source to be $\simeq 16.7$

One of the typical indicator used in the study of CV systems is the ratio between the $X$-ray and optical flux, i.e. $I=F_{X}/F_{V}$. In this case, from the above estimated $V$ magnitude the optical flux can be readily obtained to be $F_{V}\simeq 7.6\times 10^{-13}$ erg cm$^{-2}$ s$^{-1}$ and, by using the measured $X$-ray corrected flux, one obtains $I\simeq 0.23$ which is a typical value for a non-magnetic CV (for which it is expected to have $I$ in the range $10^{-3}-10$). This is consistent with the hypothesis that MACHO 104.20906.960 is a dwarf nova (non-magnetic) system (as first pointed out by \citealt{cielinski}).

Furthermore, it is well known that for dwarf novae a relation between the absolute magnitude in the V band at the outburst and the orbital period exists (see \citealt{warner} for details), i.e.
\begin{equation}
M_V^{\rm burst}=5.74-0.259 P({\rm h})~~~~~~~~{\rm for~~~~~ P({\rm h})\ut< 15\,h}.
\end{equation}
In the case of MACHO 104.20906.960, the dereddend burst apparent magnitude ($m_V^{\rm burst}$) and the outburst absolute magnitude ($M_V^{\rm burst}$, as given by the previous equation once $P(h)$ is known) read out to be $\sim 15.2$ and $\sim 5.21$ respectively, so that the distance to the source can be easily determined to be approximately $d\simeq 993$ pc. Hence, assuming isotropic emission, the $X$-ray luminosity of the source (in the $0.3-10$ keV band) can be derived as $L_{X}=4\pi d^2F_{X}$ obtaining $L_{X}\simeq 2\times 10^{31}$ erg s$^{-1}$, again typical for a dwarf nova (se e.g. \citealt{warner} and \citealt{erik}).

We emphasize that long term campaigns, as the past OGLE (\citealt{udalski1997}) and MACHO (\citealt{alcock2001}) experiments, are particularly attractive due to their capability of monitoring several million of stars for long time. These observational campaigns, among the other results, allowed the identification of several variable sources as CVs so that more studies based on an enhanced statistic may be performed.
Of course, thanks to the nature of many of these sources (interacting binary systems), one can get more information if a multi-wavelength search (spanning from the optical to the X-ray band) is done as in the case of MACHO 104.20906.960 whose nature is confirmed by {\it XMM}-Newton data. We suggest that a cross correlation search in different bands for this kind of sources is performed in the future.

\ack{
This paper is based on observations from XMM-Newton, an
ESA science mission with instruments and contributions directly funded by ESA
Member States and NASA. We are grateful to Erik Kuulkers and Maria Diaz for the interesting discussions. We are also grateful to the anonymous referee whose suggestions improved the paper.}


\end{document}